\documentclass[aps,prl,floatfix,onecolumn,groupedaddress]{revtex4-2}
\usepackage{graphicx}
\usepackage{epstopdf}
\usepackage{times}
\usepackage{amssymb,amsmath}
\usepackage{array}
\usepackage{multirow}
\usepackage{color}

\usepackage{mathtools}
\usepackage{setspace}
\usepackage{changes}
    \definechangesauthor[name={Per cusse}, color=orange]{per}

\usepackage{hyperref}
\hypersetup{hypertex=true,
    colorlinks=true,
    linkcolor=blue,
    anchorcolor=blue,
    citecolor=blue}
    
\usepackage{braket}
\usepackage{booktabs}
\usepackage{color}
\usepackage{threeparttable}
\usepackage{multirow}
\usepackage{mathtools}
\usepackage{makecell}

\makeatletter

\begin{document}

\title[]{Efficient and broadband quantum frequency comb generation in a monolithic AlGaAs-on-insulator microresonator}

\author{Xiaodong Zheng$^{1,3,\dagger}$, Xu Jing$^{1,2,\dagger}$, Chenbo Liu$^{1,\dagger}$, Yufu Li$^{1,\dagger}$, Runqiu He$^{1,3}$, Lina Xia$^{2}$, Fei Wang$^{1}$, Yuechan Kong$^{1}$, Tangsheng Chen$^{1}$, Liangliang Lu$^{1,2,3,\star}$, Jiayun Dai$^{1,\ddagger}$, Bin Niu$^{1,\ast}$}
\affiliation{
$^1$~National Key Laboratory of Solid-State Microwave Devices and Circuits, Nanjing Chip Valley Industrial Technology Institute, Nanjing Electronic Devices Institute, Nanjing, 210016, China\\
$^2$~Key Laboratory of Optoelectronic Technology of Jiangsu Province, School of Physical Science and  Technology, Nanjing Normal University, Nanjing 210023, China\\
$^3$~National Laboratory of Solid-State Microstructures, Nanjing University, Nanjing 210093, China\\
$^{\dagger}$These authors contributed equally to this work\\
$^{\star}$e-mail: lianglianglu@nju.edu.cn\\
$^{\ddagger}$e-mail: jydai2016@163.com\\
$^{\ast}$e-mail: niubin\underline{~}1@126.com
}
\date{\today}

\begin{abstract}
The exploration of photonic systems for quantum information processing has generated widespread interest in multiple cutting-edge research fields. Photonic frequency encoding stands out as an especially viable approach, given its natural alignment with established optical communication technologies, including fiber networks and wavelength-division multiplexing systems. Substantial reductions in hardware resources and improvements in quantum performance can be expected by utilizing multiple frequency modes. The integration of nonlinear photonics with microresonators provides a compelling way for generating frequency-correlated photon pairs across discrete spectral modes. Here, by leveraging the high material nonlinearity and low nonlinear loss, we demonstrate an efficient chip-scale multi-wavelength quantum light source based on AlGaAs-on-insulator, featuring a free spectral range of approximately 200 GHz at telecom wavelengths. The optimized submicron waveguide geometry provides both high effective nonlinearity ($\sim$550 m$^{-1}$W$^{-1}$) and broad generation bandwidth, producing eleven distinct wavelength pairs across a 35.2 nm bandwidth with an average spectral brightness of 2.64 GHz~mW$^{-2}$nm$^{-1}$. The generation of energy-time entanglement for each pair of frequency modes is verified through Franson interferometry, yielding an average net visibility of 93.1\%. With its exceptional optical gain and lasing capabilities, the AlGaAs-on-insulator platform developed here shows outstanding potential for realizing fully integrated, ready-to-deploy quantum photonic systems on chip.

\end{abstract}

\maketitle
\onecolumngrid
\setstretch{1.667}
\textbf{Introduction.} As a cornerstone of quantum information science, entanglement has evolved into an indispensable resource for quantum computing~\cite{OBrien2007,RevModPhys.79.135,Zhong2020}, secure communication~\cite{Gisin2007,RevModPhys.81.865}, and precision metrology~\cite{Giovannetti2011}, driving the second quantum revolution. Progress in photonic quantum information processing (QIP) fundamentally hinges on the generation and control of high-performance quantum light sources~\cite{Moody2020,Feng2020,Wang2024}. The predominant approach for room-temperature entangled photon pair generation is based on nonlinear optical interaction—either second-order ($\chi^{(2)}$) spontaneous parametric down-conversion (SPDC) or third-order ($\chi^{(3)}$) spontaneous four-wave mixing (SFWM) processes in crystalline or nanostructured materials. Currently, there is a growing trend towards implementing QIP through integrated quantum photonic circuits~\cite{Alexander2025,JWWang2018,Lull2020,Lull2021,JWWang2019,Feng2022,Luo2023}, which offer substantial advantages in miniaturization, stability, and power efficiency. Given these advantages, significant research efforts have been devoted to developing integrated photonic platforms, such as Hydex~\cite{Reimer2016, Kues2017}, silicon~\cite{Silverstone2014,JWWang2018,Bao2023}, silicon nitride~\cite{Liu2021,Larsen2025,Wen2022,Fan2023,PhysRevLett.133.083803,Ye:23}, aluminum nitride~\cite{Guo2017}, silicon carbide~\cite{Hu2024,LiJ2024} and lithium niobate-on-insulator~\cite{PhysRevLett.127.183601,Chapman2025,PhysRevLett.124.163603,PhysRevApplied.15.064059}. These platforms have successfully demonstrated essential QIP components such as heralded single-photon and entangled photon pair sources~\cite{Reimer2016,Kues2017,Silverstone2014,JWWang2018,Bao2023,Wen2022,Fan2023,PhysRevLett.133.083803,Guo2017,PhysRevLett.127.183601,Chapman2025,PhysRevLett.124.163603,PhysRevApplied.15.064059}, squeezed light generation~\cite{Larsen2025,Jia2025}, high-speed optical modulators~\cite{Xu2020,Wang2018}, high-efficiency single-photon detectors~\cite{Schuck2016,Sayem2020,Akhlaghi2015}, and quantum memories~\cite{PhysRevLett.125.260504}. These developments have substantially advanced critical quantum technologies including quantum teleportation~\cite{Llewellyn2020,feng2024chip}, quantum key distribution~\cite{Wen2022,Zhang2019,Zheng2021,Lin:25} and photonic quantum computing implementations~\cite{Paesani2019,Arrazola2021,Bao2023,Aghaee2025}. While remarkable progress has been made, assembling individual components into functional quantum systems remains non-trivial, as each material platform offers only a restricted set of capabilities.

Recently, aluminum gallium arsenide (AlGaAs) has remarkable potential for all-on-chip quantum photonic integrated circuits (QPIC) development and has emerged as a preferred material for quantum photonic applications due to its unique properties~\cite{Baboux2023,mi13070991,PhysRevLett.108.153605,JING2024,Appas2021}. For example, with precise aluminum alloying, the material exhibits a wide optical transparency window and ultralow two-photon absorption at the telecom wavelength. Its direct bandgap facilitates the monolithic integration of wavelength-tunable pump lasers~\cite{PhysRevLett.112.183901}, while its non-centrosymmetric crystalline symmetry enables the simultaneous exploitation of both $\chi^{(2)}$ and $\chi^{(3)}$ effects~\cite{Baboux2023}. High-speed electro-optic modulators~\cite{10.1117/12.2296173} and superconducting-nanowire single photon detectors (SNSPDs)~\cite{10.1063/1.3657518} have been integrated with GaAs/AlGaAs waveguides. These features, combined with the recent development of low-loss AlGaAs-on-insulator (AlGaAsOI) photonic integration~\cite{Ottaviano2016,Chang2018,Chang2020}, herald a transformative opportunity and powerful platform for scalable nonlinear applications, such as efficient optical frequency comb generation~\cite{Chang2020,Pu2016} and octave-spanning frequency conversion~\cite{Stassen2019}. Recent breakthroughs have extended the desired nonlinear properties of this platform into the quantum regime~\cite{PRXQuantum.2.010337}, demonstrating an AlGaAsOI microresonator array capable of generating multiplexed entangled frequency combs~\cite{Pang2025}. The discrete, equally-spaced frequency modes emerging from these resonators enable exponential scaling of the system's dimensionality with minimal resource overhead, while being fully compatible with standard telecom multiplexing and modulation schemes. Meanwhile, frequency encoding offers a resource-efficient approach to accessing high-dimensional Hilbert spaces within a single spatial mode, paving the way for scalable QIP.  When operating at telecom wavelengths (near 1550 nm), frequency bins can be controlled using commercial off-the-shelf devices~\cite{Kues2017}, including electro-optic phase modulators (EOMs) and programmable filters (PFs). Studies have demonstrated that combining two EOMs with a PF enables arbitrary qubit transformations~\cite{PhysRevLett.120.030502,PhysRevLett.125.120503}. The inherent reconfigurability of such quantum frequency processors supports diverse applications in quantum information science~\cite{Poolad2020,LuH2023,Karthik2025}. Moreover, the multi-wavelength quantum light sources are attractive due to their compatibility with standard optical fiber infrastructure and their ability to perform wavelength-based routing, and have proven to be an indispensable component to fully-connected multiple-user quantum networks~\cite{Wengerowsky2018,Joshi2020,Wen2022}. Building upon these breakthroughs, developing high-flux multi-wavelength quantum light sources can offer pivotal advantages for large-scale QIP. However, the number of frequency channels, governed by the microring's cavity length (and thus its free spectral range, FSR), is fundamentally constrained by a trade-off with the achievable pair generation rate~\cite{PRXQuantum.2.010337}.

Here, we strategically designed the microring’s FSR to align with the 200 GHz International Telecommunication Union (ITU) dense wavelength-division multiplexing (DWDM) grid, thereby balancing these competing requirements. The waveguide cross-sectional dimension is engineered to enhance optical confinement, thereby achieving an effective nonlinearity of 550 m$^{-1}$ W$^{-1}$, while maintaining a broadband operational bandwidth. The combination of high nonlinearity and larger microring radius enables multi-wavelength quantum light generation, producing 11 high-quality correlated photon pairs across a range of 35.2 nm, with a free spectral range (FSR) of about 200 GHz in the telecom C-band. The 11 correlated photon pairs exhibit an average spectral brightness of 2.64 GHz~mW$^{-2}$nm$^{-1}$, and enable the generation of energy-time entangled photons with of 93.1\% net visibility in the Franson interferometer measurement. These findings establish AlGaAsOI microring resonators as a promising platform for future quantum network implementations.

\textbf{Device fabrication and characterization.} Fig. 1(a) shows our AlGaAsOI microring resonator that is fabricated with a 400 nm thick Al$_{0.2}$Ga$_{0.8}$As photonic layer bonded on a 2.0 $\mu$m silica-on-silicon wafer. The microring resonator has a radius of 58.6 $\mu$m. A pulley-style coupler is employed, in which the bus waveguide adiabatically approaches the microring, ensuring high coupling ideality. The widths of the bus and microring waveguide are 0.5 $\mu$m and 0.6 $\mu$m, respectively, with a 0.38 $\mu$m gap between them. A segment of the transmission spectrum of the microring resonator from 1527 to 1565 nm is shown in Fig. 1(b) with an averaged FSR of 202.8 GHz, which is adapted to commonly used International Telecommunication Union’s (ITU) dense wavelength-division multiplexing (DWDM) filters. The averaged extinction ratio (ER) is 11.2 dB, with the extracted intrinsic and loaded quality (Q) factors of 2.2 $\times$ 10$^5$ (Q$_i$) and 7.8 $\times$ 10$^4$ (Q$_L$), respectively. Fig. 1(c) displays the transmission spectrum at the resonant wavelength of 1546.58 nm (C39) where the pump locates with a Q$_L$ of 8.3 $\times$ 10$^4$. Anomalous and near-zero dispersion properties are crucial for SFWM phase matching in the device. As shown in Fig. 1(d), we  measure the dispersions of the microring resonator and fit the data to the integrated dispersion: $D_{int}(\mu) = \omega_{\mu} - (\omega_0 + D_1\mu) = 1/2 D_2 \mu^2 + O(\mu^3)$, where $\mu$ is the relative mode number, $\omega_0/2\pi$ is the pump resonance’s frequency, $D_1/2\pi$ is the FSR, and $D_2/2\pi$ denotes the group velocity dispersion (GVD) parameter. The fitted $D_2/2\pi$ is 32.4 MHz, indicating an anomalous dispersion of -1.12 $\times$ 10$^{-24}$ s$^2$ m$^{-1}$ in a wide spectrum. The detailed device fabrication and characterization are presented in Supplementary Information.

\begin{figure*}[!htbp]
\begin{center}
    \includegraphics[width=0.9\textwidth]{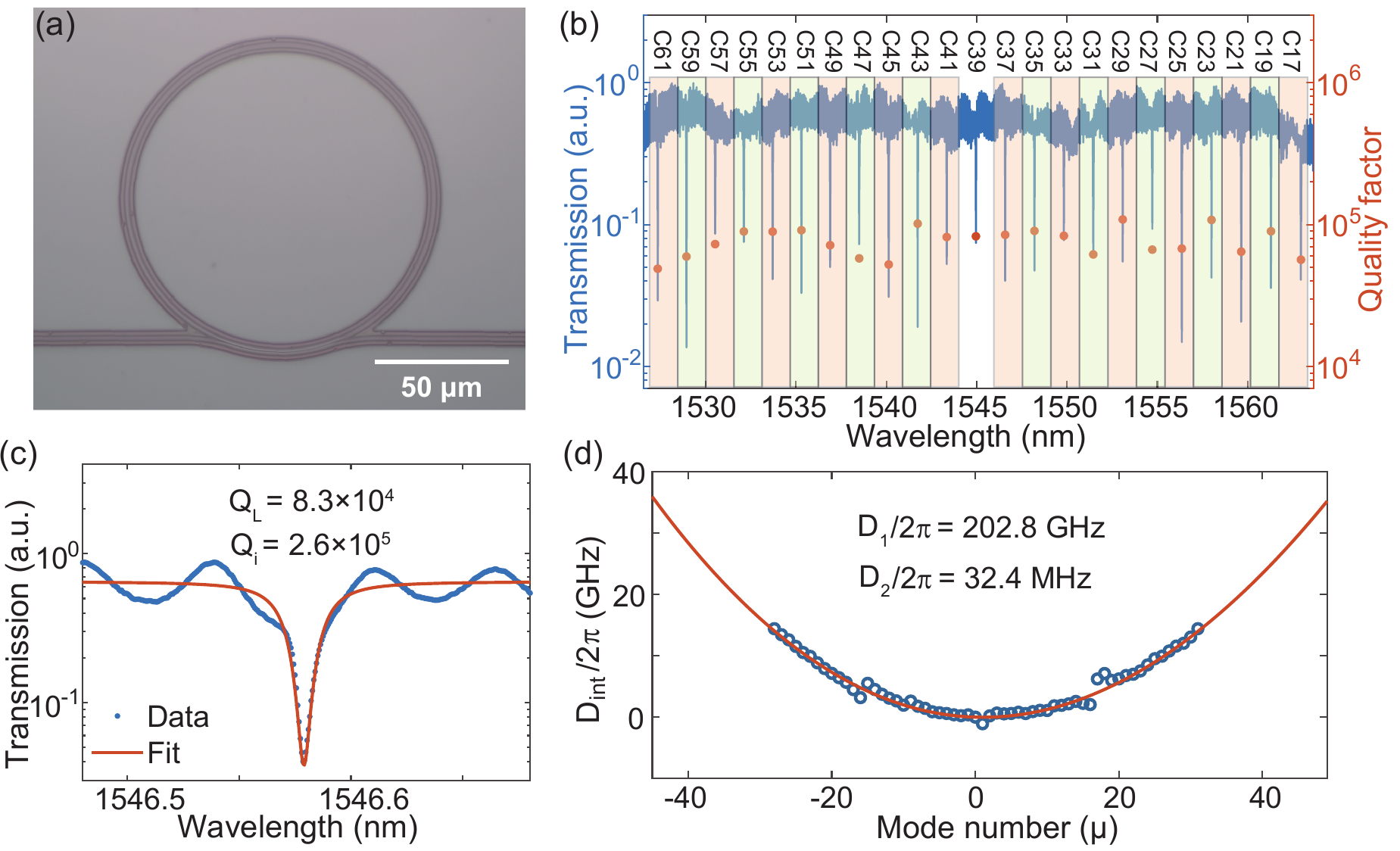}
    \caption{(a) Optical microscopy image of the AlGaAsOI microring resonator. (b) Measured transmission spectrum and loaded Q factors from 1527 to 1565 nm with an averaged free spectral range (FSR) of about 200 GHz and Q factor of 7.8 $\times$ 10$^4$. (c) Measured and Lorentz fitted resonance dip around 1546 nm (C39) where the pump locates with a loaded, intrinsic and external Q factors of 8.3 $\times$ 10$^4$, 2.6 $\times$ 10$^5$ and 1.2 $\times$ 10$^5$, respectively. (d) Integrated dispersion $D_{int}$ extracted from the transmission and calculated based on the resonator wavelengths, indicating the second-order GVD of -1.12 $\times$ 10$^{-24}$ s$^2$ m$^{-1}$.} 
\end{center}
\end{figure*}

\textbf{Multi-wavelength correlated photon-pair generation.} The scheme for generating and characterizing multi-wavelength correlated photon pairs is depicted in Fig. 2. Fig. 2(a) illustrates the experimental setup for the generation of photon pairs. The pump light emitted from a continuous-wave (CW) tunable laser at a wavelength of 1546.12 nm corresponding to ITU channel of C39. A fiber polarization controller (PC) is used to control the polarization state of the pump light. Pump sideband noise and Raman noise generated in the fiber pigtails are suppressed using cascaded high-isolation ($\ge$ 120 dB) DWDMs at C39 with a 20-cm-long pigtail followed by a 15-cm-long lens tapered fiber before the chip. The pump light is coupled in and out the chip by lens tapered fibers mounted on high-precision servo motors. The total insertion loss of the chip is about -18.0 dB, including both input-output coupling loss and propagation loss in the chip. Correlated photon pairs are generated through SFWM in the microring resonator, where two pump photons are annihilated to create signal and idler photon pairs. These photon pairs exhibit discrete comb-like spectral modes resulting from the resonator's natural spectral filtering characteristics. At the output of the chip, a 100 GHz DWDM filter centered at C39 serves as a band-stop filter to remove the residual pump light. The correlated photon pairs are selected by a programmable wavelength-selective switch (WSS) as shown in Fig. 2(b).  The photons are detected by two InGaAs avalanche single photon detectors (APDs), which operate with 25\% detection efficiency, 1.6 kHz dark count rates, and 10 $\mu$s dead time. The detection signals from APDs are sent to a logic circuit to record photon arrival time. The total loss of the signal and idle photons is about -49.0 dB.

\begin{figure*}[htbp]
\begin{center}
    \includegraphics[width=0.9\textwidth]{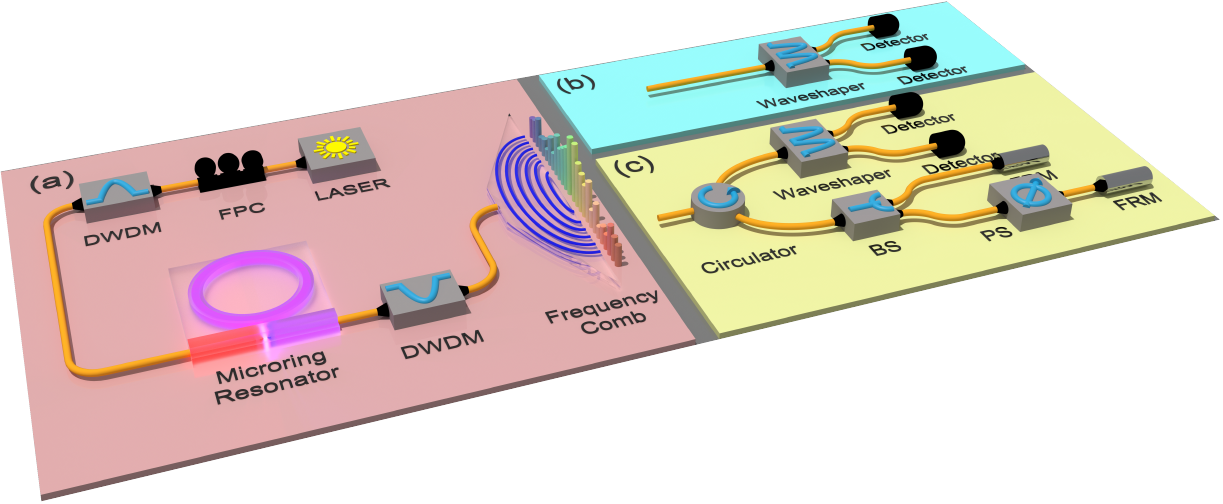}
    \caption{Experimental setups for characterizing the multi-wavelength quantum light source. (a) Multi-wavelength correlated photon-pair generation. (b) Correlation properties. (c) Energy-time entanglement with two-photon interference.}
\end{center}
\end{figure*}

\begin{figure*}[htbp]
\begin{center}
    \includegraphics[width=0.9\textwidth]{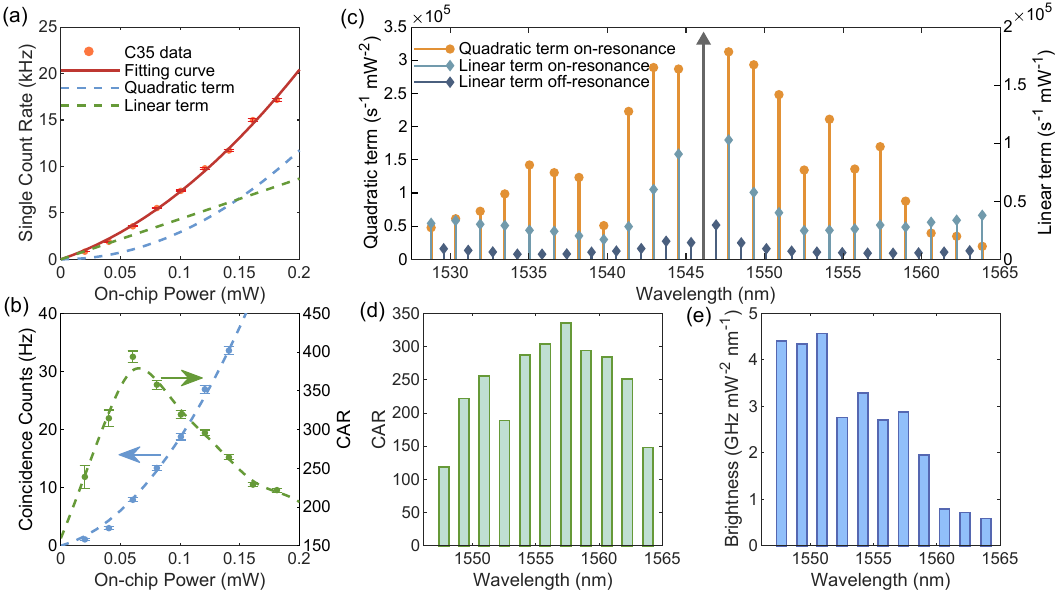}
    \caption{Measured results of generated multi-wavelength correlated photon pairs. (a) Single count rates of idler (C43) photon with different on-chip power and fitting results. (b) Coincidence count rate and the calculated CAR of signal (C35) and idler (C43) photons with different on-chip power. (c) Spectra of the correlated photons and noise photons in the on-resonance and off-resonance cases from 1527 nm to 1565 nm. (d) CAR of multi-wavelength correlated photons. (e) Spectral brightness of multi-wavelength correlated photons.}
\end{center}
\end{figure*}
In Fig. 3(a), we select the idler mode in C43 as an example to indicate the dependence of single count rate on on-chip pump power (red dots). The data are fitted to the quadratic function $aP^2+bP+c$, where the quadratic term $aP^2$ is the contribution of correlated photons (blue dash line), the linear term $bP$ corresponds to the contribution of noise photons (green dash line), and the constant term $c$ accounts for dark counts. The error bars are calculated as standard deviations from the Poisson distribution of the raw counts. Experimentally, we obtain $a$ = 2.93 $\times$ 10$^5$ s$^{-1}$ mW$^{-2}$, $b$ = 4.35 $\times$ 10$^4$ s$^{-1}$ mW$^{-1}$. To further assess the quality of the generated photon pairs, Fig. 3(b) presents the measured coincidence count rates and coincidence-to-accidental ratio (CAR) with a 0.5 ns coincidence window, recorded for signal and idler photons at 1542.94 nm (C35) and 1549.32 nm (C43), respectively. We observe a CAR of 394 with an 8 Hz coincidence rate at 0.06 mW on-chip pump power.  After compensating for the total transmission losses from chip to detectors, we calculate an on-chip photon pair generation rate of 65.1 MHz mW$^{-2}$. We extract the third-order nonlinear optical coefficient $\gamma \approx$ 550 m$^{-1}$W$^{-1}$ (see Supplementary Information for details), more than two orders of magnitude higher than silicon nitride~\cite{Fan2023}, highlighting the superior nonlinear performance of AlGaAsOI. To demonstrate the multi-wavelength operation of the quantum light source, we perform spectral characterization of both correlated photons and noise photons by programming the WSS as a tunable band-pass filter. The WSS is scanned across the full C-band from 1528.77 nm (C61) to 1563.86 nm (C17) in 100 GHz increments, with the spectral range constrained by the WSS operational bandwidth. As shown in Fig. 3(c), single count rates at different resonance wavelengths with different on-chip pump power are recorded. The quadratic (blue dots) and linear (red dots) terms for all resonant wavelength channels are extracted. The black dots represent the linear terms from noise photons in the microring's off-resonance case, which are significantly smaller than the on-resonance linear terms. This difference confirms the generation of spontaneous Raman scattering noise photons within the microring resonator under resonant conditions~\cite{Reimer2016,PhysRevA.75.023803}. This is mainly due to the defects from the lattice mismatching between the AlGaAs layer and the SiO$_2$ buffer layer. Fig. 3(d) shows the CAR for various wavelength pairs at a fixed on-chip pump power of 0.18 mW. All measured CAR values exceed 118, demonstrating excellent signal-to-noise performance across the entire spectral range. We also display the spectral brightness of 11 wavelength-paired resonances spanning a 35.2 nm bandwidth in Fig. 3(e). The average brightness of 2.64 GHz~mW$^{-2}$nm$^{-1}$ demonstrates the quantum light source's dual achievement of both broadband operation and high brightness.

\begin{figure*}[!h]
\begin{center}
    \includegraphics[width=0.8\textwidth]{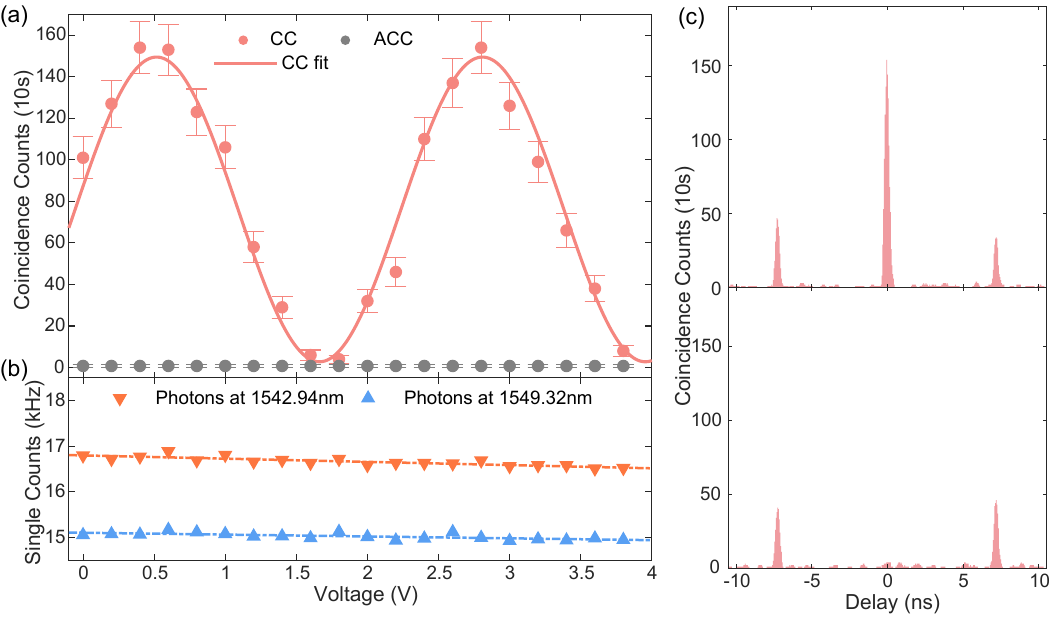}
   \caption{Two-photon interference and energy-time entanglement. (a) Coincidence counts (red dots) and accidental coincidence counts (grey dots) of signal (C35) and idler (C43) photons with different phase shifter’s voltages. (b) single count rate of signal (C35) and idler (C43) photons with different phase shifter’s voltages. The sinusoidal fit of shows a raw visibility of 0.9636(217) without background subtraction. (c) Two-photon correlation histograms of constructive (top) and destructive (bottom) two-photon interference.}
\end{center}
\end{figure*}

\textbf{Energy-time entanglement and two-photon interference.} The generated multi-pair correlated photons are expected to exhibit energy-time entanglement~\cite{PhysRevLett.82.2594}, which offer distinct advantages for long-distance distribution, as they exhibit strong resistance to decoherence and immunity to random polarization rotation in optical fibers~\cite{PhysRevLett.93.180502,Qi2021,Huang2025,Wengerowsky2018,Joshi2020,Wen2022,Liu2022}. The energy-time entanglement can be measured through a Franson-type two-photon interference experiment. As shown in Fig. 2(c), we use a folded Franson interferometer to interfere two temporally separated photons. Signal and idler photons are both injected into the fiber circulator and then are separated into two paths by a 50/50 fiber beam splitter. In the lower path, a voltage driven fiber phase shifter is used to control the phase $\phi_i + \phi_s$ of signal and idler photons. At each path end, a Faraday rotation mirror is connected to return light with polarization rotated 90$^\circ$ relative to the input, which can minimize alterations to the polarization state induced by thermal and mechanical perturbations in the fiber. The interferometer path length difference is set to $\Delta t$ = 7 ns, which is longer than the single-photon coherence time to avoid single-photon interference. The two paths are then recombined and interfere. Signal and idler photons can either take a short ($S$) or a long ($L$) path in the interferometer. We select the case when both photons take the same path, which gives the energy-time entanglement state $\ket{\Psi} = 1/\sqrt{2}(\ket{SS} +e^{i(\phi_i + \phi_s)}\ket{LL})$. To suppress phase fluctuation in the fiber due to ambient temperature fluctuation, we place the interferometer in a heat-insulated container. The measured two-photon interference of (C35) and idler (C43) photons is shown in Fig. 4(a). The coincidence counts oscillate periodically with the phase shifter’s voltages, while single count rates of signal and idler photons are constant indicating no single-photon interference in our measurement. Using sinusoidal fitting, we analyze coincidence counts both with and without accidental subtraction, obtaining a raw visibility of V$_{raw}$ = 0.9636 $\pm$ 0.0217 and a net visibility of  V$_{net}$ = 0.9731 $\pm$ 0.0188 at 0.18 mW on-chip pump power. The maximum values of $S$ parameter $S_{max}$ is 2.725 $\pm$ 0.061 (2.752 $\pm$ 0.053), which give a violation of the Clauser–Horne–Shimony–Holt (CHSH) inequality by more than 12 (14) standard deviations. In our analysis, we determine the background level by averaging 10 randomly selected bin values from non-peak spectral regions. The coincidence histograms for $\phi_i + \phi_s$ = 0 and $\pi$/2 are shown in Fig. 4(c), which represent the constructive and destructive two-photon interference, respectively. The measured energy-time entanglement properties of other wavelength-paired photon pairs are given in Table \ref{tab:1}. All eleven wavelength channel pairs exhibit interference visibilities exceeding 86.0 \%, confirming the high-quality energy-time entanglement characteristics of this multi-wavelength quantum light source.

\begin{table}[htbp]
\caption{Results of the correlated photon pairs at different wavelengths.}
\label{tab:1}
\centering
\begin{threeparttable}
\resizebox{0.9\linewidth}{!}{
\begin{tabular}{ c  c  c  c  c }
\toprule
$\lambda_s \& \lambda_i$ (nm) & \makecell{Coherence time\\(ps)} & \makecell{Brightness\\(GHz~mW$^{-2}$nm$^{-1}$)} & Net visibility (\%) & $S_{max}$\\
\midrule
1544.53 $\&$ 1547.72 & 61.87 & 4.41 & 99.36 $\pm$ 1.00 & 2.810 $\pm$ 0.028 \\ 
1542.94 $\&$ 1549.32 & 87.60 & 4.34 & 97.31 $\pm$ 1.88 & 2.752 $\pm$ 0.053\\ 
1541.35 $\&$ 1550.92 & 69.33 & 4.57 & 93.74 $\pm$ 3.11 & 2.651 $\pm$ 0.088\\ 
1539.77 $\&$ 1552.52 & 73.26 & 2.76 & 89.71 $\pm$ 4.42 & 2.537 $\pm$ 0.125\\ 
1538.19 $\&$ 1554.13 & 58.56 & 3.28 & 93.81 $\pm$ 3.12 & 2.653 $\pm$ 0.088\\ 
1536.61 $\&$ 1555.75 & 49.08 & 2.71 & 93.12 $\pm$ 3.32 & 2.634 $\pm$ 0.094\\ 
1535.04 $\&$ 1557.36 & 67.45 & 2.88 & 90.98 $\pm$ 3.50 & 2.573 $\pm$ 0.099\\ 
1533.47 $\&$ 1558.98 & 35.58 & 1.95 & 97.65 $\pm$ 1.89 & 2.762 $\pm$ 0.053\\ 
1531.90 $\&$ 1560.61 & 84.06 & 0.79 & 92.39 $\pm$ 4.35 & 2.613 $\pm$ 0.123\\ 
1530.33 $\&$ 1562.23 & 68.95 & 0.72 & 89.62 $\pm$ 4.36 & 2.535 $\pm$ 0.123\\ 
1528.77 $\&$ 1563.86 & 86.36 & 0.59 & 86.23 $\pm$ 5.31 & 2.439 $\pm$ 0.150\\ 
\bottomrule
\end{tabular}
}
\end{threeparttable}
\end{table}

\textbf{Discussion and conclusion.} In summary, we have demonstrated a high-quality chip-scale multiwavelength quantum light source with an AlGaAsOI microring resonator. With its high refractive index contrast and exceptional effective nonlinearity of 550 m$^{-1}$W$^{-1}$, this platform enables highly-efficient generation of 11 wavelength-channel photon pairs across a wavelength range of 35.2 nm. The average spectral brightness reaches 2.64 GHz~mW$^{-2}$nm$^{-1}$. Energy-time entanglement for the multi-wavelength quantum light source has been confirmed with an averaged net two-photon interference visibility of 93.1 \%. Inspired by the recently developed high-performance microresonator-based quantum light sources~\cite{PhysRevLett.133.083803,PRXQuantum.2.010337,PhysRevLett.82.2594,LiJ2024,Placke2024,LiB2025}, we would like to suggest a further avenue for achieving both higher generation rates and increased spectral multiplexing capacity. In theory, the photon-pair generation rate is scaled with $Q^3/R^2$, where $Q$ and $R$ is the quality factor and radius of microresonator respectively. Therefore, the photon-pair generation rate can be enhanced by either enhancing the quality factor of microresonator or reducing the cavity length. The key to reach a higher quality factor is optimizing the fabrication process to reduce the scattering and bending losses. Photoresist reflow~\cite{Gyorgy2014,Xie2020}, dry-etch process optimization~\cite{Kim2022} and surface passivation~\cite{Guha:17} have been used to effectively reduce losses. Considering the emission processes of the photon pairs, an optimized over-coupled with a ratio between external quality factor (Q$_e$) and intrinsic quality factor (Q$_i$) of 3/5 should be satisfied~\cite{Fan2023,Guo:18} by designing the gap width between the bus waveguide and the microring. The multi-wavelength property, depending on the FSR of the microring, can be improved by multiplexing arrays of small-radius microresonators~\cite{Pang2025}. It is also desirable to make a near-zero and flat anomalous dispersion over a large wavelength range with sophisticated phase-matching by designing the thickness and width of the waveguide (see Supplementary Information for more details), such that the broadband, high-generation-rate, and multi-wavelength channel quantum light sources can be achieved. Furthermore, our findings reveal new possibilities for implementing frequency-domain quantum processing in integrated photonics ~\cite{Lukens:17,PhysRevA.109.013505,Clementi2023}. The exceptional potential for large-scale integration and versatile integration capabilities of AlGaAs position it as a key enabler for next-generation quantum information technologies. 

\section{Acknowledgements}
\noindent
This work was supported by the National Natural Science Foundation of China (12274233), the Natural Science Foundation of Jiangsu Province (SBK20250402321), the Innovation Program for Quantum Science and Technology (2021ZD0300700), Nanjing Electronic Devices Institute (2311N051), and the Postgraduate Research $\&$  Practice Innovation Program of Jiangsu Province (SJCX25$\_$0691).\noindent
\noindent
\bibliographystyle{naturemag}
\bibliography{reference}

\clearpage
\newpage

\section{Supplementary information}
\section{Device fabrication and characterization}\label{sec:one}
\begin{figure*}[htbp]
\begin{center}
    \includegraphics[width=0.9\textwidth]{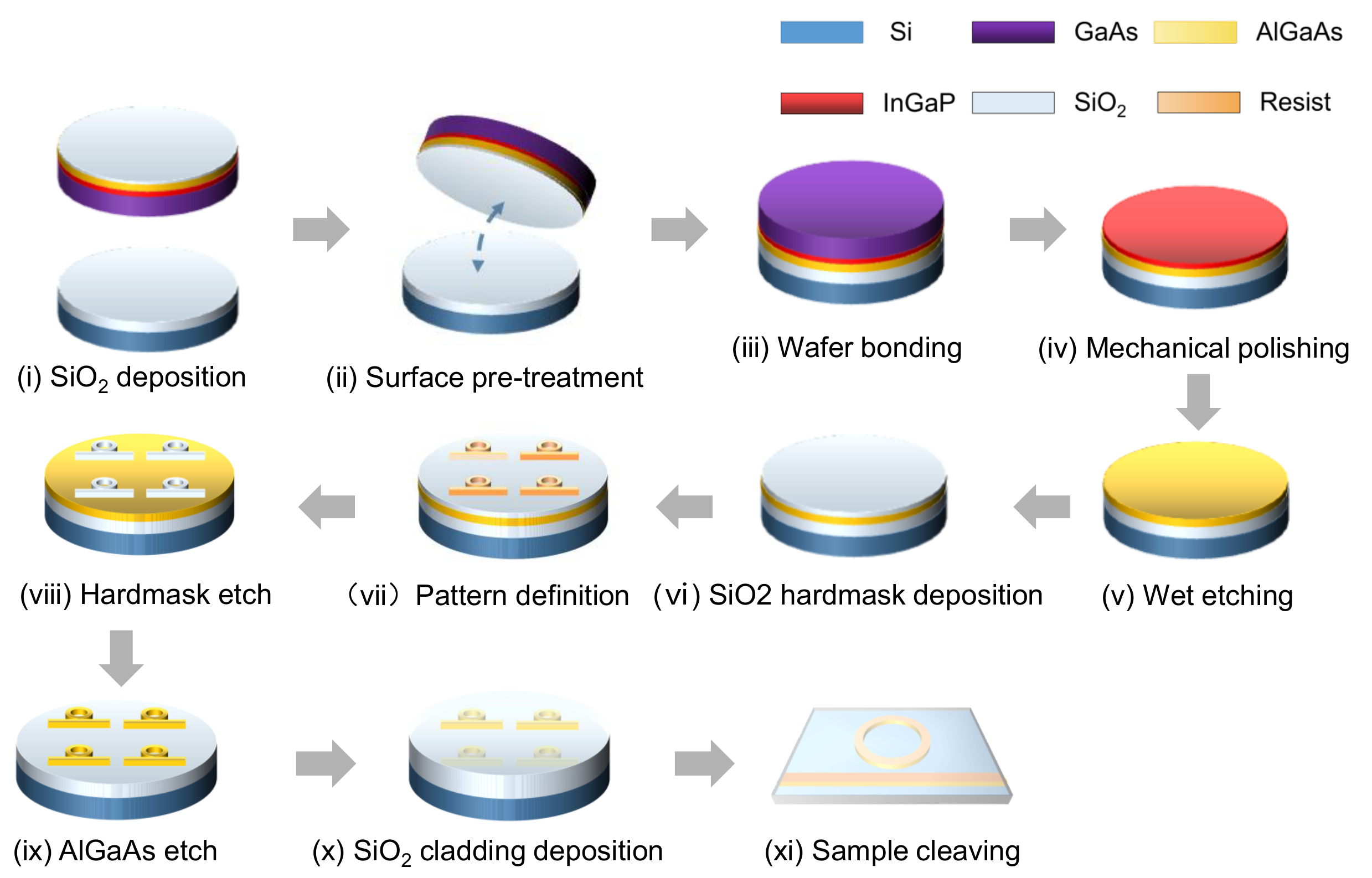}
    \caption{\label{Fig5}The process flow of AlGaAsOI device fabrication.}
\end{center}
\end{figure*}

The device fabrication process is illustrated in Fig. \ref{Fig5}. The thin 400 nm thick Al$_{0.2}$Ga$_{0.8}$As with sacrificial layers (InGaP/ GaAs/ InGaP) is epitaxially grown using molecular beam epitaxy (MBE) on a 4-inch GaAs wafer (100 surface). A cladding layer, SiO$_2$ with a thickness of 2.0 $\mu$m, is deposited on the prepared AlGaAs epitaxial stack surface through plasma enhanced chemical vapor deposition (PECVD). The SiO$_2$ cladded AlGaAs wafer is bonded to a thermally oxidized silicon carrier wafer with a 10 nm thick AlN adhesive layer at high vacuum environment. Then the bonded sample is placed in an oven at 120 $\mathrm{^\circ C}$ for 12 hours under 1 MPa pressure to enhance the bonding strength. Afterward, mechanical polishing was applied to lap the GaAs substrate thickness down to 25 $\mu$m. Subsequently, the residual GaAs is slowly removed using a mixture of citric acid and hydrogen peroxide ($\mathrm{C_6H_8O_7 : H_2O_2 = 4 : 1}$) at a much slower etching rate. Following the removal of the GaAs substrate, the first etching stop layer, InGaP, is removed using phosphoric acid and hydrogen chloride ($\mathrm{H_3PO_4 : HCl = 4 : 1}$). The intermediate GaAs layer and the last InGaP etching stop layer are removed using the same citric acid/hydrogen peroxide mixture and the phosphoric acid /hydrogen chloride mixture, respectively, leaving only the Al$_{0.2}$Ga$_{0.8}$As photonic layer on the carrier wafer. At this point, the fabrication of AlGaAsOI is complete.

\begin{figure*}[htbp]
\begin{center}
    \includegraphics[width=0.8\textwidth]{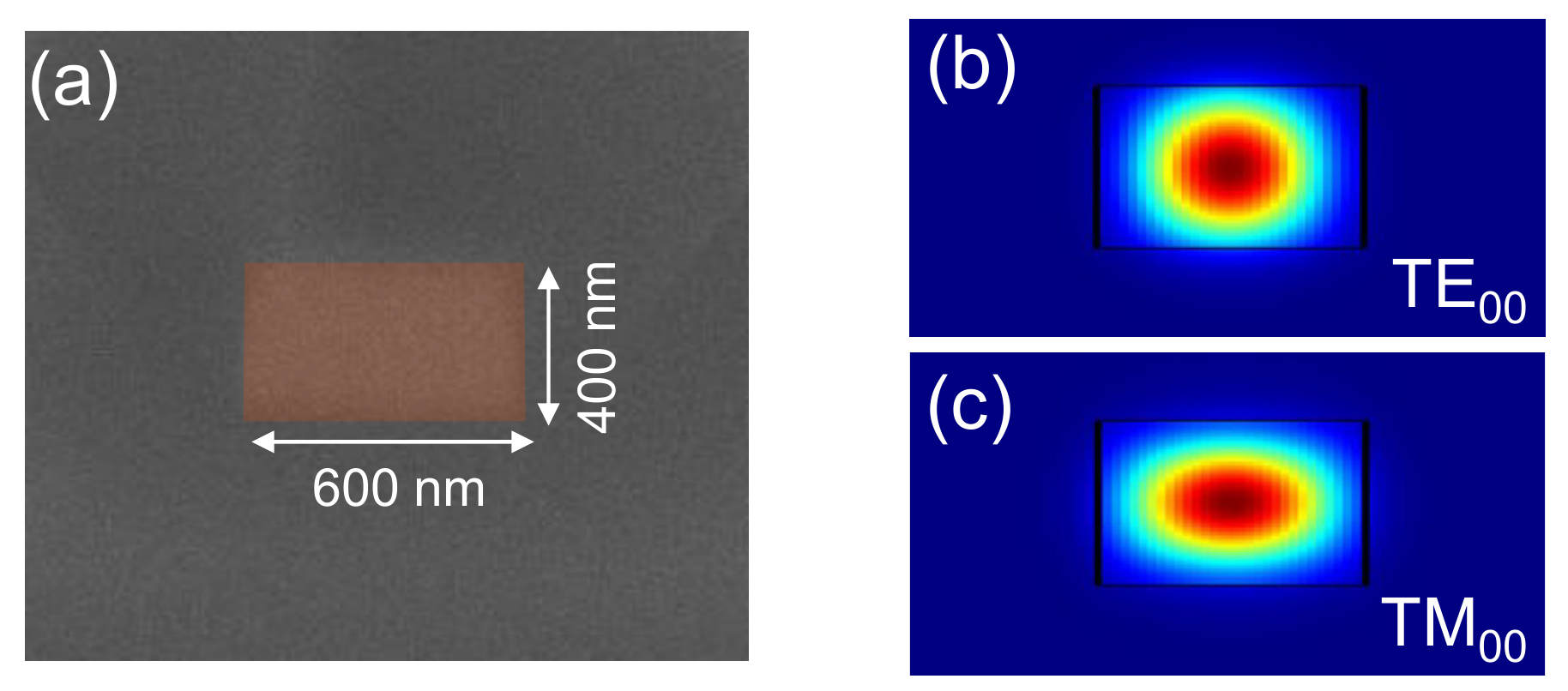}
    \caption{(a) The SEM image of waveguide cross section. (b) and (c) are the simulated mode profiles of TE$_{00}$ and TM$_{00}$.}\label{Fig. S2}
\end{center}
\end{figure*}

After substrate removal, a 150 nm SiO$_2$ layer is deposited on surface of AlGaAsOI as a hardmask. Electron-beam lithography (EBL) and a positive electron-beam resist AR6200.13 are used to define our device patterns. After exposure and development of the resist, a thermal reflow process is applied to the wafer at 152 $\mathrm{^\circ C}$ on a hotplate for 1.5 minutes. An inductively coupled plasma -reactive ion etching (ICP-RIE) system is used to dry etch the SiO$_2$ hardmask with CHF$_3$ gas. After removal of residual resist, we use Cl$_2$ /N$_2$ gases to pattern the photonic device structure on the AlGaAs layer. A 2.0 $\mu$m SiO$_2$ cladding layer is deposited on top of the dry-etched AlGaAsOI sample with the PECVD. Finally, the sample is thinned to 200 $\mu$m and cleaved at each end of the bus waveguide nanotapers. In our work, a AlGaAs microring with a radius of 58.6 $\mu$m and a waveguide cross-section of 0.6 $\mu$m $\times$ 0.4 $\mu$m, which is shown in Fig. \ref{Fig. S2}(a). The simulated mode profiles of TE$_{00}$ and TM$_{00}$ are shown in Fig. \ref{Fig. S2}(b) and (c), demonstrating strong light confinement.

\begin{figure*}[htbp]
\begin{center}
    \includegraphics[width=1\textwidth]{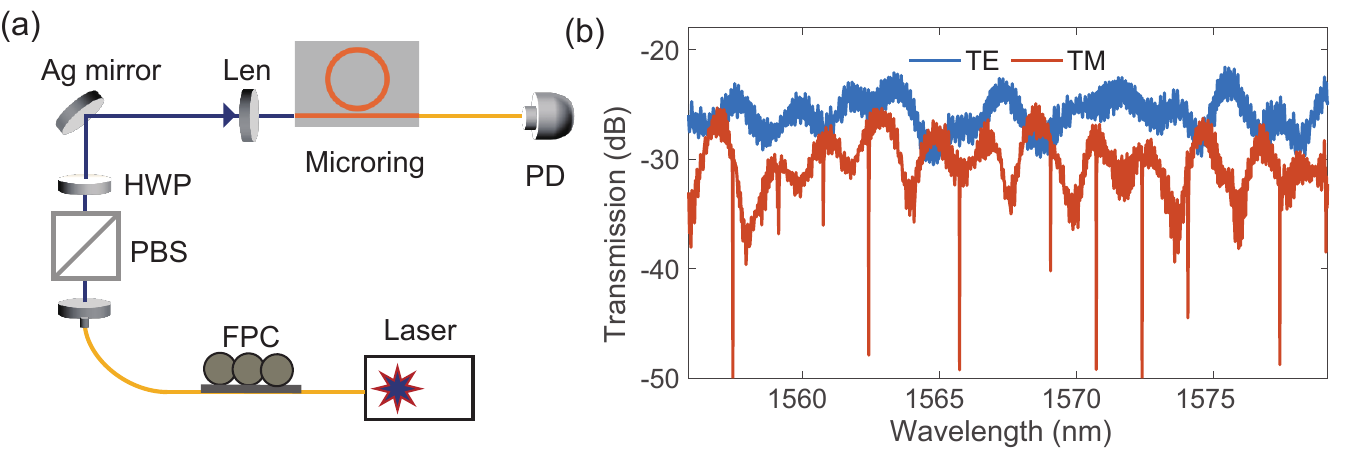}
    \caption{(a) The experimental setup of transmission spectrum measurement. (b) The measured transmission spectrum for TE and TM modes.}\label{Fig. S3}
\end{center}
\end{figure*}

The transmission spectrum of the microring resonator can be used to distinguish the transmission mode of light in the waveguide in addition to extracting parameters such as quality factor and extinction ratio. The TE$_{00}$ and TM$_{00}$ mode transmission properties of the waveguide can be generally characterized and distinguished by the free spectral range (FSR) due to the difference of group index. In our device, only one polarization mode’s resonance transmission spectrum was measured. We verify the polarization mode of transmission spectrum by employing the setup as shown in Fig. \ref{Fig. S3}(a). The light from a tunable continuous-wave (CW) tunable laser is connected to a fiber polarization controller and a collimator. A polarization beamsplitter (PBS) cube and a half wave plate are used to identify the polarization of incident light. The light is coupled into the AlGaAs nanotaper waveguide using a lens. The transmission spectrum of TE and TM incident light are shown in Fig. \ref{Fig. S3}(b). The transmission spectrum shows a distinct resonance dip for TM mode, whereas the TE mode exhibits negligible resonant effects. This is attributed to the high coupling loss and Fabry-Perot interference at the chip interface, which obscure the resonance detection for TE modes. On the other hand, the TE$_{00}$ mode exhibits lower coupling efficiency between the bus waveguide and the ring resonator compared to the TM$_{00}$ mode. See Section \ref{sec:five} for more details. 

\section{Analysis of experimental data}\label{sec:two}
In this section, the extraction of photon pair generation rate ($R_{PG}$), two-photon interference visibility ($V$) and the violation of Clauser-Horne-Shimony-Holt (CHSH) Bell inequality from the photon counts are analyzed in detail.

The photon pair generation rate ($R_{PG}$) is a key figure of merit for quantum light sources. It quantifies the number of correlated photon pairs produced per unit time and per unit pump power via a nonlinear process. Experimentally, $R_{PG}$ can be extracted from the measured single-photon count rates of the signal and idler photons under varying pump powers. The detected single-photon count rate ($N_{s(i)}$) (where subscripts s and i denote signal and idler photons, respectively) originates not only from spontaneous four-wave mixing (SFWM), but also includes contributions from spontaneous Raman scattering and detector dark counts. This total measured count rate can be expressed as:
    \begin{gather}
        N_s=\eta_s R_{PG} P^2+R_{RS,s} P+DC_s, \\
        N_i=\eta_i R_{PG} P^2+R_{RS,i} P+DC_i, 
    \end{gather}
    \begin{figure*}[htbp]
\begin{center}
    \includegraphics[width=1\textwidth]{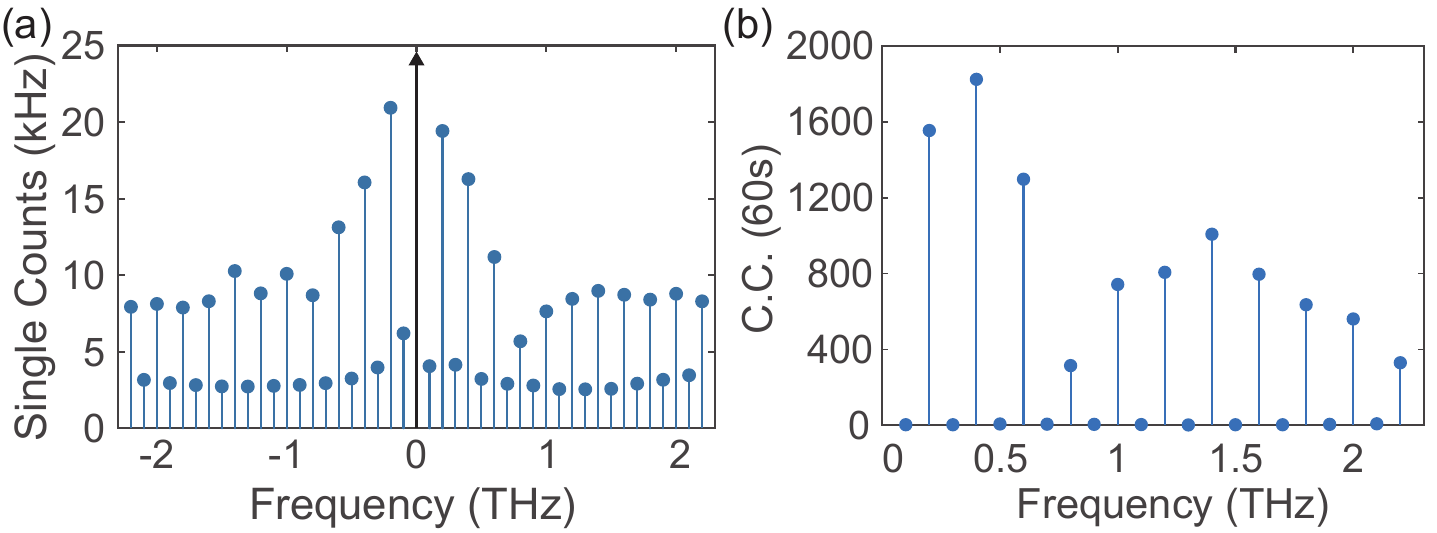}
    \caption{The single-photon count rate (a) and coincidence counts (b) for 11 channels of correlated photon pairs and noise photons at on-chip pump power of 0.16 mW.}\label{Fig.S4}
\end{center}
\end{figure*}
where $P$ is the on-chip pump power, $\eta_{s\left(i\right)}$ is the total collection efficiency for the signal (idler) photon, encompassing the chip insertion loss, filtering loss, and detection loss. $R_{RS}$ represents the Raman‑scattering‑induced noise, which may originate from the input/output fibers, waveguides, and the microring resonator. ${DC}_{s(i)}$ is the dark count rate of the detector. We denote the quadratic term coefficient $\eta_{s(i)}R_{PG}$ as $R_{s\left(i\right)}$. After correcting for the detector dead time ($\tau_{dead}$), the measured single photon count rate can be written as:
    \begin{equation}
        {SC}_{s\left(i\right)}=N_{s\left(i\right)}\frac{1}{1+\tau_{dead}N_{s\left(i\right)}}.\label{EQ.11}
    \end{equation}
The measured coincidence count ($CC$) and the accidental coincidence count ($ACC$) can be expressed, respectively, as: 
\begin{gather}
    CC=\eta_s\eta_iR_{PG}P^2+ACC,\\
    ACC={SC}_s{SC}_it_{CC},
\end{gather}
where $t_{CC}$ is the temporal width of the coincidence window. The quadratic term coefficient of $CC$ is denoted as $R_{CC}$. Consequently, the photon-pair generation rate can be determined by combining the coefficients extracted from the single-channel and coincidence measurements:
    \begin{equation}
        R_{PG}=\frac{R_sR_i}{R_{CC}}.\label{EQ.12}
    \end{equation}
Finally, the spectral brightness ($B$)--defined as the number of photon pairs generated per second, per unit pump power, and per unit bandwidth--is obtained by normalizing $R_{PG}$ by the effective spectral bandwidth of the source. The coincidence-to-accidental ratio (CAR) a key metric for quantifying signal purity, is given by: $CAR=(CC-ACC)/ACC$.
\begin{table}[htbp]
\caption{The $R_{PG}$ and $B$ of the correlated photon pairs at different wavelengths.}
\label{tab:1}
\centering
\begin{threeparttable}
\resizebox{0.9\linewidth}{!}{
\begin{tabular}{ c  c  c  c }
\toprule
$\lambda_s \& \lambda_i$ (nm) & Coherence time (ps) & $R_{PG}$ (MHz~mW$^{-2}$) & B (GHz~mW$^{-2}$nm$^{-1}$)\\
\midrule
1544.53 $\&$ 1547.72 & 61.87 & 66.1 & 4.41  \\ 
1542.94 $\&$ 1549.32 & 87.60 & 65.1 & 4.34 \\ 
1541.35 $\&$ 1550.92 & 69.33 & 68.5 & 4.57 \\ 
1539.77 $\&$ 1552.52 & 73.26 & 41.5 & 2.76 \\ 
1538.19 $\&$ 1554.13 & 58.56 & 49.3 & 3.28 \\ 
1536.61 $\&$ 1555.75 & 49.08 & 40.6 & 2.71 \\ 
1535.04 $\&$ 1557.36 & 67.45 & 43.1 & 2.88 \\ 
1533.47 $\&$ 1558.98 & 35.58 & 29.3 & 1.95 \\ 
1531.90 $\&$ 1560.61 & 84.06 & 11.9 & 0.79 \\ 
1530.33 $\&$ 1562.23 & 68.95 & 10.7 & 0.72 \\ 
1528.77 $\&$ 1563.86 & 86.36 & 8.8 & 0.59 \\ 
\bottomrule
\end{tabular}
}
\end{threeparttable}
\end{table}
To obtain the values of $R_{PG}$ and $B$ for photon pairs across different wavelength channels, we performed measurements by programming the wavelength selective switch (WSS, Waveshaper) as a tunable band-pass filter and recording counts at various on-chip pump powers.The WSS was scanned across the full C-band from 1528.77 nm (ITU channel C61) to 1563.86 nm (C17) in 100 GHz increments; the scan range was limited by the operational bandwidth of the WSS. In Fig. \ref{Fig.S4}, we provide the single-photon count rate and coincidence counts for 11 channels of correlated photon pairs and noise photons at on-chip pump power of 0.16 mW. Measurements were also taken at other power levels. For each wavelength channel pair, the quadratic coefficients of the single-photon count rate and the coincidence counts were extracted by fitting the measured count rates as a function of on-chip pump power (see the “Multi-wavelength correlated photon-pair generation” section in the main text for details). Using the method described above, the photon-pair generation rate $R_{PG}$ and spectral brightness $B$ results for different wavelength channel correlated photon pairs were listed in Table \ref{tab:1}.

The energy-time entanglement state $\ket{\Psi} = 1/\sqrt{2}(\ket{SS} +e^{i(\phi_i + \phi_s)}\ket{LL})$ is generated using a Franson interferometer~\cite{PhysRevLett.82.2594,PhysRevLett.93.180502}. The coincidence count $CC$ for the central peak (corresponding to $\ket{SL}$ ($\ket{LS}$) terms, as shown in Fig. 4(c)) exhibits a sinusoidal modulation as a function of the relative phase. This is plotted in Fig. 4(a) and can be described by the function:
    \begin{equation}
        CC \sim 1 + ij V cos(\phi_i + \phi_s), \label{EQ.1}
    \end{equation}
where $i,j = \pm 1$ and $V$ is visibility of the interference fringes (which can, in principle, reach a maximum value of 1). Here, $\phi_i$ and $\phi_s$ represent the phases of the idler and signal photon, respectively. The correlation coefficient $E(\phi_i, \phi_s)$ is defined as:
    \begin{equation}
        E(\phi_i, \phi_s) = \frac{\sum_(i,j) ij CC}{\sum_(i,j) CC}. \label{EQ.2}
    \end{equation}
Substituting Eq. \ref{EQ.1} into Eq. \ref{EQ.2}, the correlation coefficient simplifies to:
    \begin{equation}
        E(\phi_i, \phi_s) = V cos(\phi_i + \phi_s). \label{EQ.3}
    \end{equation}
The Clauser-Horne-Shimony-Holt (CHSH) Bell inequality~\cite{CHSH} is given by:
    \begin{equation}
        S = \lvert E(\phi_i, \phi_s) + E(\phi_i, \phi^{'}_{s}) + E(\phi{'}_i, \phi_s) - E(\phi{'}_i, \phi{'}_s)\rvert \leq 2. \label{EQ.4}
    \end{equation}
\begin{figure*}[htbp]
\begin{center}
    \includegraphics[width=1\textwidth]{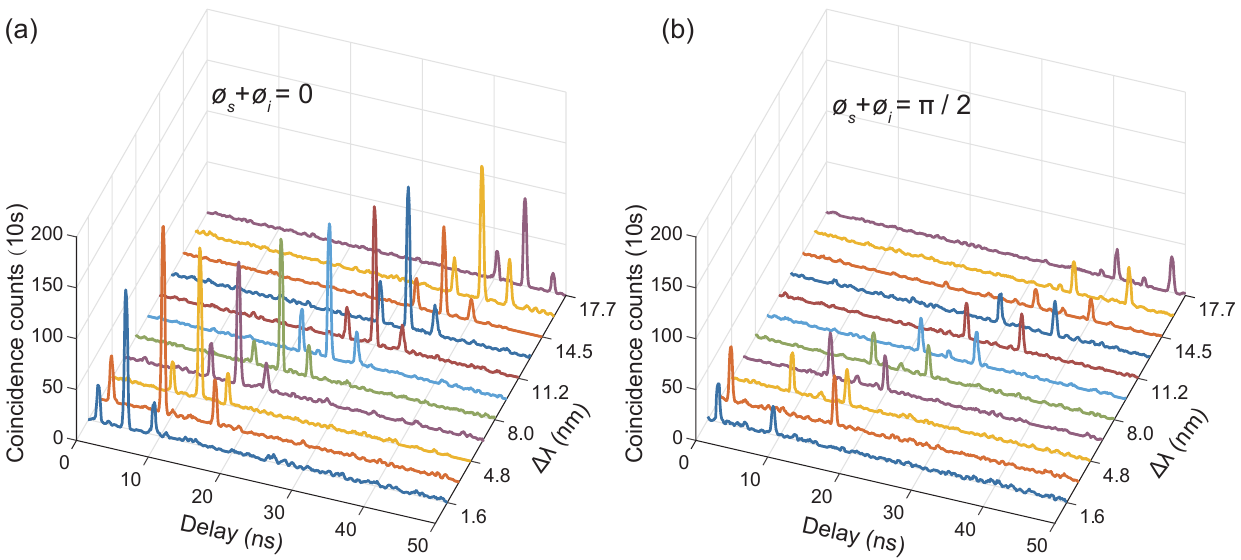}
    \caption{The coincidence histograms of 11 pairs scanned ITU channels for $\phi_i + \phi_s$ = 0 and $\pi$/2.}\label{Fig. S6}
\end{center}
\end{figure*}
\begin{table}[htbp]
\caption{Results of Franson interference of the correlated photon pairs at different wavelengths.}
\label{tab:2}
\centering
\begin{threeparttable}
\begin{tabular}{ c  c  c  c  c }
\toprule
$\lambda_s \& \lambda_i$ (nm) & $V_{raw}$ (\%) & $V_{net}$ (\%) & $S_{max}$  &  $n_{\sigma}$\\
\midrule
1544.53 $\&$ 1547.72 & 96.64 $\pm$ 2.26  & 99.36$\pm$1.00  & 2.810$\pm$0.028 &29 Std. \\ 
1542.94 $\&$ 1549.32 & 96.36$\pm$2.17 &  97.31$\pm$1.88 & 2.752$\pm$0.053  & 14 Std.\\ 
1541.35 $\&$ 1550.92 & 92.85$\pm$3.30 & 93.74$\pm$3.11  & 2.651$\pm$0.088 & 7 Std..\\ 
1539.77 $\&$ 1552.52 & 88.63$\pm$4.60  & 89.71$\pm$4.42  & 2.537$\pm$0.125 & 4 Std.\\ 
1538.19 $\&$ 1554.13 & 92.71$\pm$3.35 & 93.81$\pm$3.12  & 2.653$\pm$0.088 & 7 Std.\\ 
1536.61 $\&$ 1555.75 & 92.07$\pm$3.53 &  93.12$\pm$3.32  & 2.634$\pm$0.094 & 7 Std.\\ 
1535.04 $\&$ 1557.36 & 89.46$\pm$3.73 & 90.98$\pm$3.50  & 2.573$\pm$0.099 &6 Std.\\ 
1533.47 $\&$ 1558.98 & 96.40$\pm$2.32 &  97.65$\pm$1.89  & 2.762$\pm$0.053 &14 Std.\\ 
1531.90 $\&$ 1560.61 & 91.09$\pm$4.66 &  92.39$\pm$4.35 & 2.613$\pm$0.123  & 5 Std.\\ 
1530.33 $\&$ 1562.23 & 88.55$\pm$4.54 &  89.62$\pm$4.36  & 2.535$\pm$0.123 & 4 Std.\\ 
1528.77 $\&$ 1563.86 & 85.00$\pm$5.49 &  86.23$\pm$5.31 & 2.439$\pm$0.150 &  3 Std.\\ 
\bottomrule
\end{tabular}
\end{threeparttable}
\end{table}
Quantum mechanics predicts that $S$ reaches a maximum value of $S_{max} = 2\sqrt{2}$ for the choice of phases $\phi_i = 0$, $\phi{'}_i = \pi/2$, $\phi_s = \pi/4$, and $\phi^{'}_{s} = -\pi/4$. Furthermore, it has been shown that when the correlation function follows the sinusoidal form of Eq. \ref{EQ.3} and the system exhibits rotational invariance, the bound in Eq. \ref{EQ.4} can be expressed as~\cite{RevModPhys.86.419}:
    \begin{equation}
        S_{max} = 2\sqrt{2} V \leq 2, \label{EQ.5}
    \end{equation}
Therefore, a visibility $V > 1/\sqrt{2}$ directly implies a violation of the CHSH Bell inequality.
The visibility $V$ can be directly obtained from the interference curve data (e.g., Fig. 4(a)) using the relation:
    \begin{equation}
        V = \frac{CC_{max}-CC_{min}}{CC_{max}+CC_{min}}, \label{EQ.6}
    \end{equation}
where $CC_{max}$ and $CC_{min}$ are the maximum and minimum coincidence counts, respectively. The extent of violation of the CHSH Bell inequality, expressed in terms of standard deviations $n_{\sigma}$, is defined as:
    \begin{equation}
        n_{\sigma} = \frac{S_{max} - 2}{\sigma_{S}}, \label{EQ.7}
    \end{equation}
where $\sigma_{S}$ is the standard deviation of the $S_{max}$. Through error propagation, the $\sigma_{S}$ can be calculated from the measurement data as follows:
    \begin{equation}
        \sigma_{S} = 2\sqrt{2} \sigma_{V} = 2\sqrt{2} \sqrt{(\frac{2 CC_{min}}{(CC_{max}+CC_{min})^2}\sigma_{CC_{max}})^2 + (\frac{-2 CC_{max}}{(CC_{max}+CC_{min})^2}\sigma_{CC_{min}})^2}, \label{EQ.8}
    \end{equation}
    \begin{figure*}[htbp]
\begin{center}
    \includegraphics[width=\textwidth]{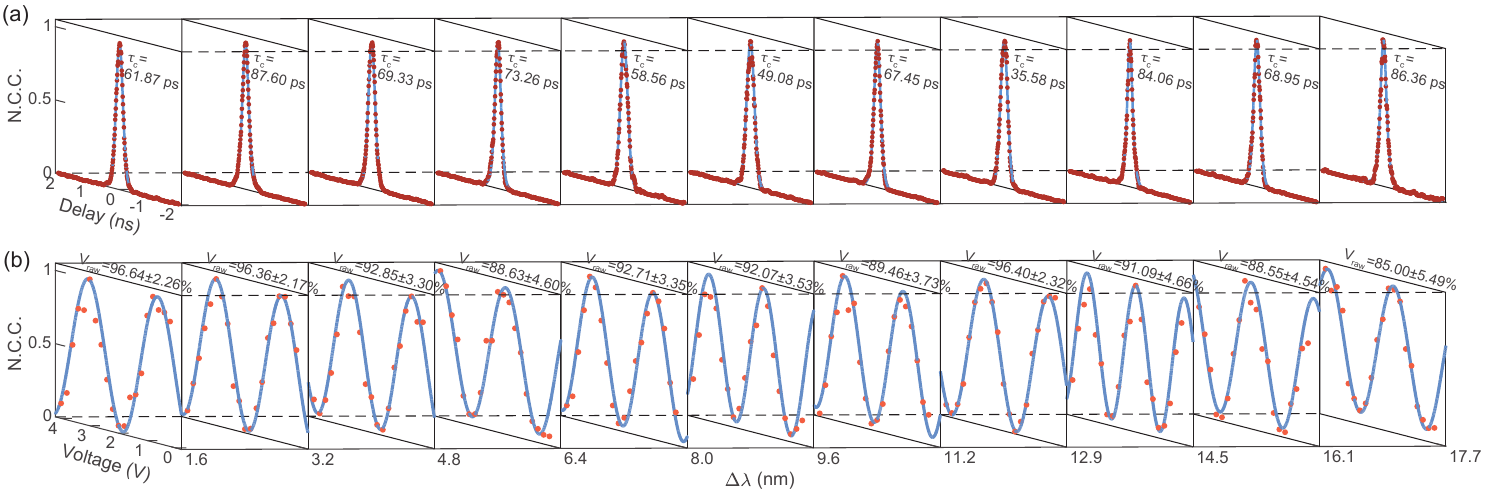}
    \caption{(a) The fitting results of the histograms of different wavelength pairs’ coincidence counts. (b) The fitting results of different wavelength pairs’ Franson interference curves.}\label{Fig. S7}
\end{center}
\end{figure*}
where $\sigma_{CC}$ representing the standard error of the coincidence count. Since the coincidence counts follow Poisson statistics, $\sigma_{CC} = \sqrt{CC}$. Thus, the violation of the CHSH Bell inequality can be rigorously determined from the measured visibility or directly from the coincidence counts. The coincidence histograms of 11 pairs scanned ITU channels for $\phi_i + \phi_s$ = 0 and $\pi$/2 are shown in Fig. \ref{Fig. S6}.
Then we calculate the raw/net visibility ($V_{raw}$/$V_{net}$), $S_{max}$ and violation of the CHSH Bell inequality ($n_{\sigma}$) for all 11 pairs scanned ITU channels. The results are listed at Table \ref{tab:2}. It should be noted that the raw visibility is calculated directly from the detection data, and the net visibility is obtained by the data which removes the background level. In our analysis, we determine the background level by averaging 10 randomly selected bin values from non-peak spectral regions.

The photon-pair generation is evidenced by the two-photon correlation histogram shown in Fig. \ref{Fig. S7}(a). The coincidence count (CC) is obtained by taking a window with a center at zero relative delay and a width of 2.5 ns from the histogram of coincidence counts. The accidental coincidence count (ACC) is the average of the counts in all other 2.5 ns windows except the zero delay in the histogram. The measured coincidence count curve is the convolution of the signal /idler temporal correlation function and the response function of the whole detection system, which can be written as $1 / \tau_c~e^{-\left| \Delta t \right| /\tau_c} \ast e^{⁡-(\Delta t)^2/\sigma^2}$, where $\Delta t$ is the time delay, $\tau_c$ is the coherence time of the single photons and $\sigma$ is the time jitter of the detection system. After considering the time jitter, we obtained an average coherence time of 67.5 ps. In Fig. \ref{Fig. S7}(b), we give the Franson interference curves of eleven entangled photon pairs. The recorded raw interference visibilities are listed, demonstrating an average visibility of 91.8 \%. 

\section{Losses of different components}\label{sec:three}
\begin{table}[htbp]
\caption{Losses of different components.}
\label{tab:3}
\centering
\begin{threeparttable}
\resizebox{0.4\linewidth}{!}{
\begin{tabular}{ c  c}
\toprule
Component & Loss\\
\midrule
Chip insertion loss &  9.0 dB per facet  \\ 
Pre-filter (DWDMs) & 1.4 dB \\ 
Post-filter (DWDMs) & 1.6 dB \\ 
Waveshaper & 4.5 dB \\ 
Interferometer &  2.5 dB\\ 
Detector &  6.0 dB\\ 
\bottomrule
\end{tabular}
}
\end{threeparttable}
\end{table}
The losses of each component in the experimental setup are summarized in the Table \ref{tab:3}. The primary cause of the high chip insertion loss is the absence of end-face polishing processes. Using chemical mechanical polishing (CMP) technology can effectively reduce this loss. Furthermore, defining the facet via dry etching~\cite{10.1063/5.0098984} is also a method of reducing chip insertion loss. As for the loss of other components in the experimental setup, the equipment's inherent losses are unavoidable. However, connection losses between devices (e.g., filters comprising multiple DWDMs arranged in a cascade) can be minimized through fiber splicing.


\section{Nonlinear optical coefficient}\label{sec:four}

The nonlinear optical coefficient ($\gamma$) is an important parameter relevant in evaluating the nonlinear performance of a waveguide. The AlGaAsOI waveguide's cross-sectional dimensions are engineered at the submicron scale to enhance light confinement, thereby increasing the device's effective nonlinearity. In the SFWM process, the generation rate ($N$) of correlated photons is given by Eq. \ref{EQ.9}\cite{Fan2023,Guo:18},
    \begin{equation}
        N = \frac{32\gamma^2 v^4_g P^2_p Q^8}{\omega^3_0 L^2 Q^5_e} sinc^2(\frac{L \Delta \kappa}{2}),\label{EQ.9} 
    \end{equation}
where $\gamma$, $v_g$, $P_p$, $Q$, $\omega_0$, $L$, $Q_e$ and $\Delta \kappa$ denote the nonlinear optical coefficient, group velocity, pump power, total (loaded) Q-factor, angular frequency on-resonance, roundtrip length of the cavity, external quality factor, and the phase mismatch, respectively. The group velocity and Q-factor can be extracted from the transmission spectrum. The phase-matching condition $sinc^2(\frac{L \Delta \kappa}{2})=1$ is fulfilled when the angular frequency of correlated photons is near the angular frequency of pump light. Thus, the nonlinear optical coefficient ($\gamma$) can be calculated from the measured correlated-photon count rate. Our measurements yield a nonlinear optical coefficient of $\gamma \approx$ 550 m$^{-1}$W$^{-1}$.

\section{Coupling and dispersion optimization}\label{sec:five}
\begin{figure*}[htbp]
\begin{center}
    \includegraphics[width=0.9\textwidth]{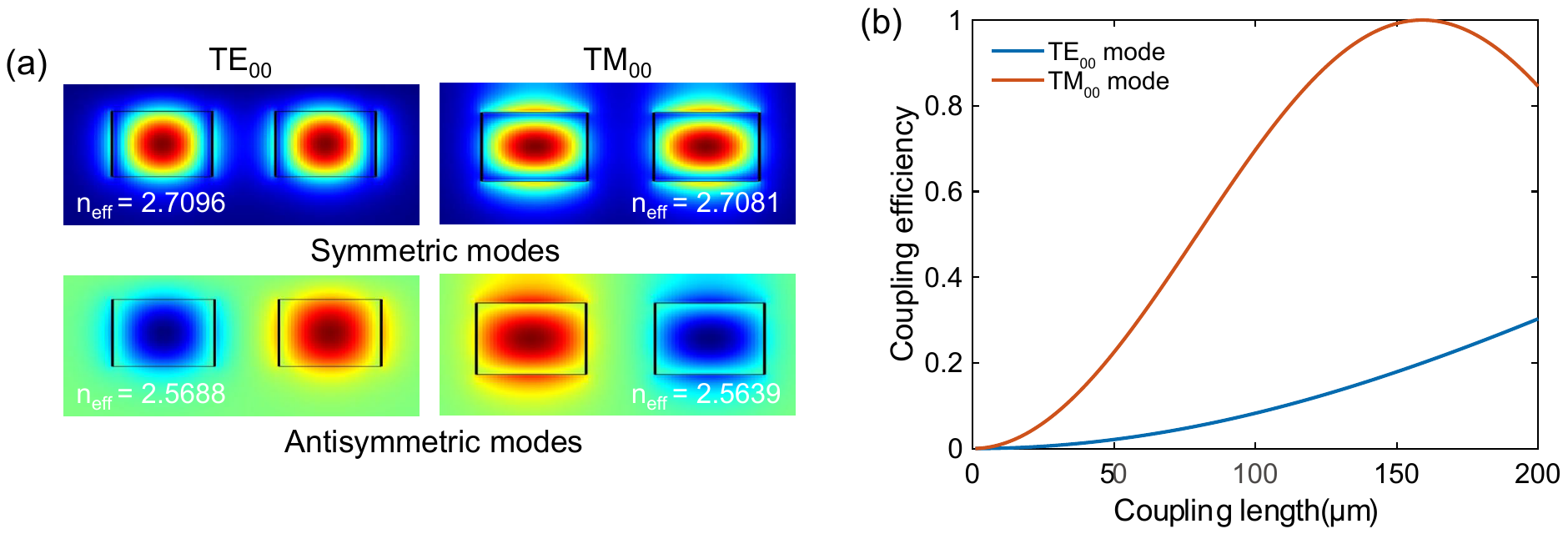}
    \caption{(a) The simulated mode profiles of symmetric and antisymmetric modes. (b) The calculated coupling efficiency for TE$_{00}$ and TM$_{00}$ as a function of the coupling length.}\label{Fig. S8}
\end{center}
\end{figure*}
\begin{figure*}[htbp]
\begin{center}
    \includegraphics[width=0.9\textwidth]{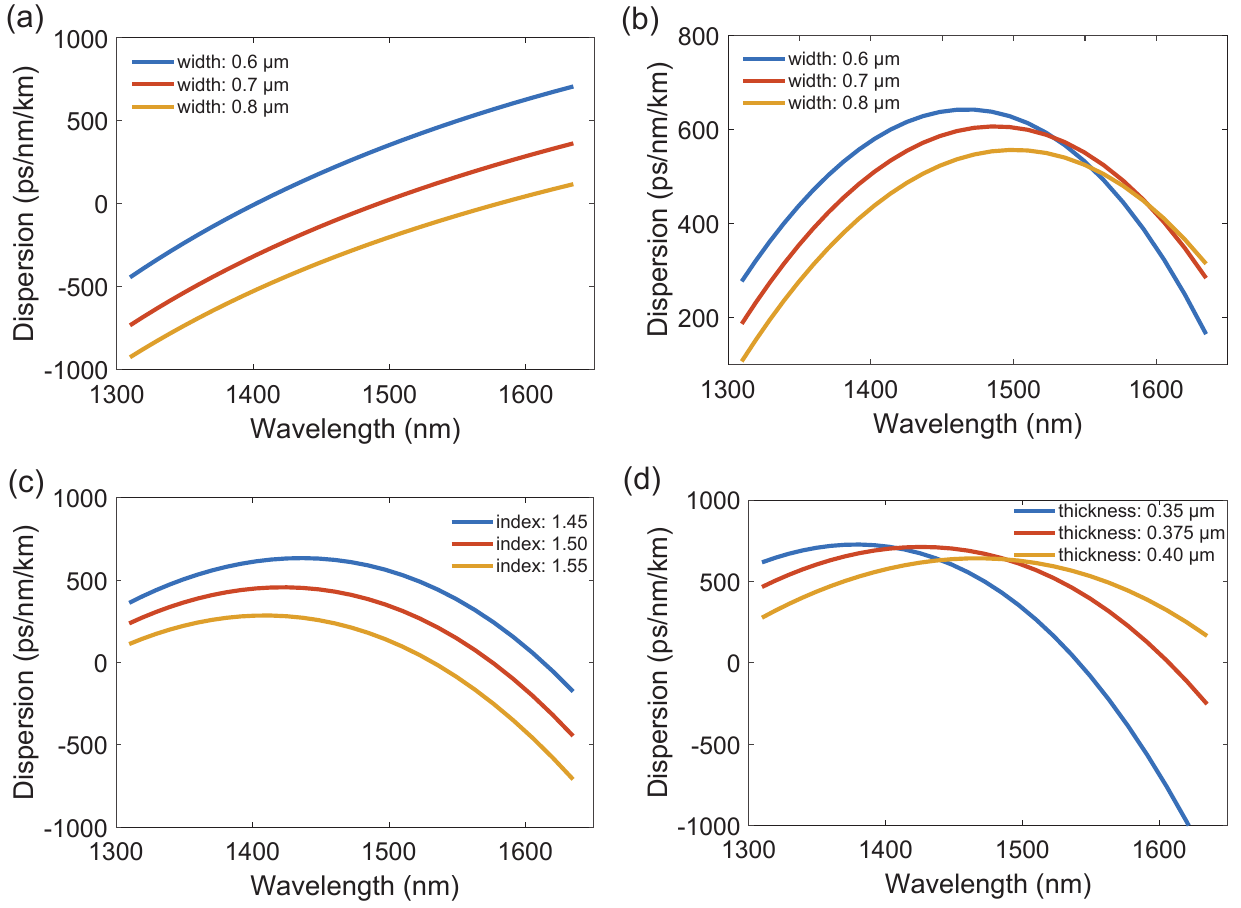}
    \caption{The TE (a) and TM (b) mode dispersion of a 400 nm thick AlGaAs waveguide changes for various waveguide widths. (c) The TM$_{00}$ mode dispersion of an AlGaAs waveguide with a cross section of 0.5 $\mu$m $\times$ 0.4 $\mu$m changes for various refractive index of the cladding layer. (d) The TM$_{00}$ mode dispersion of an AlGaAs waveguide with a 600 nm width changes for various height of the waveguide.}\label{Fig. S9}
\end{center}
\end{figure*}

The large index contrast of AlGaAsOI enables the design of waveguides with high mode confinement which enhances the nonlinear interaction and allows for more compact structures to be created. However, this tight confinement can sometimes make it difficult to efficiently couple light from one waveguide to another in a short distance. Thus, in the design of ring resonators, point coupler may not transfer enough power into the ring for efficient quantum light generation. The design of pulley couplers, where the waveguide and ring bend together for a portion of the ring, can allow for more power to be transferred into the ring. The coupling coefficient can be calculated from the effective index of the symmetric and antisymmetric modes using eigenmode expansion (EME) mode solver. As shown in Fig. \ref{Fig. S8}(a), we simulate the symmetric and antisymmetric modes of TE$_{00}$ and TM$_{00}$ with waveguide cross-section of 0.6 $\mu$m $\times$ 0.4 $\mu$m. The coupling coefficient ($C$) can be calculated as
    \begin{equation}
        C = \frac{\pi \Delta n_{neff}}{\lambda}, 
    \end{equation}
where $\Delta n_{neff}$ and $\lambda$ denote the effective index of the symmetric and antisymmetric modes and wavelength of propagating light. We can find that the TM$_{00}$ mode  exhibits a coupling coefficient approximately five times larger than that of the TE$_{00}$ mode. Therefore, longer coupling length is needed for TE$_{00}$ mode. Figure \ref{Fig. S8}(b) presents the calculated coupling efficiency, defined as the fraction of power coupled into the ring, for both TE$_{00}$ and TM$_{00}$ modes as a function of coupling length.

In our device, the GVD parameter $D_2$ is slightly larger, which limit the wavelength number of the light source. Broad spectrum can be achieved when the waveguides have close to zero dispersion at the pump wavelength. Fig. \ref{Fig. S9}(a) and (b) shows how the dispersion of a 400 nm thick AlGaAs waveguide changes for TE and TM modes with various waveguide widths. At 1550 nm the dispersion of TE$_{00}$ mode is approximately 0 ps/nm/km for an 800 nm wide waveguide. However, the dispersion of TM$_{00}$ mode is close to 500 ps/nm/km for an 800 nm wide waveguide. To achieve zero dispersion for the TM$_{00}$ mode,  an alternative approach involves modifying the refractive index of the cladding layer. As shown in Fig. \ref{Fig. S9}(c), the dispersion of TM$_{00}$ mode is close to zero when the cladding SiO$_2$ layer’s index is about 1.50, which is achievable by changing the deposition process parameters. Alternatively, the condition can also be achieved by reducing the waveguide height from 400 nm to 375 nm through dry etching techniques or by precisely controlling the thickness of epitaxially grown AlGaAs layers (Fig. \ref{Fig. S9}(d)).

\section{Comparison of quantum light sources}\label{sec:six}
To show the performance of our multi-wavelength quantum light source, we compare the performance of quantum light sources with different material platforms as shown in Table \ref{tab:4}.

\begin{table}[htbp]
\caption{Comparison of microring quantum light sources.}
\label{tab:4}
\centering
\begin{threeparttable}
\resizebox{1\linewidth}{!}{
\begin{tabular}{ c  c  c  c  c  c  c  c  c  c  }
\toprule
Reference& Material & Radius ($\mu$m) & Quality factor & $R_{PG}$ (Hz~mW$^{-2}$) & B (Hz~mW$^{-2}$nm$^{-1}$)& CAR & Visibility (\%)& Wavelength channel& Year\\
\midrule
\cite{Reimer:14}& Hydex & 135 & 1.4$\times$10$^6$ & 2.86$\sim$3.46$\times$10$^5$ & 2.55$\sim$3.09$\times$10$^8$& 14& -& 5& 2014  \\ 
\cite{ramelow2015}&SiN & 115 & 2$\times$10$^6$ & 3.9$\times$10$^6$ &8.1$\times$10$^9$ & -& 90& 1& 2015 \\ 
\cite{Grassani:15}&SOI & 10 & 1.5$\times$10$^4$ & $\sim$6.2$\times$10$^6$  & $\sim$6$\times$10$^7$& 132& 95& 1& 2015\\ 
 \cite{Ma:17}&SOI & 10 & 9.2$\times$10$^4$ & 1.49$\times$10$^8$  &8.87$\times$10$^9$ & 12105& 95.9& 1& 2017\\ 
\cite{Fujiwara:17} &SOI & 10  &2$\times$10$^4$& $\sim$4$\times$10$^7$  & $\sim$5.2$\times$10$^8$  & 350& 82& 1& 2017\\ 
\cite{Kumar}&InP & 24 & 4.2$\times$10$^4$ & $\sim$1.45$\times$10$^8$  &$\sim$3.9$\times$10$^9$ & 277& 78& 1& 2019\\ 
\cite{Samara:19}&SiN & - & 4.6$\times$10$^5$ & 5.2$\times$10$^5$ &$\sim$3.1$\times$10$^8$ & 495& 98.3& 1& 2019\\ 
\cite{RN248}&AlGaAs & 13.9 & 1.2$\times$10$^6$ & 2$\times$10$^{10}$ & 2.5$\times$10$^{13}$& 2697& 97.1&1 &2021 \\ 
\cite{Samara_2021}&SiN & - & 4.4$\times$10$^5$ & 5.5$\times$10$^4$ &1.4$\times$10$^7$ &- & 98.7& 1& 2021\\ 
\cite{Fan2023}&SiN  & 113 & 1$\times$10$^6$ & 1.15$\times$10$^6$ &1.2$\times$10$^9$& 1243& 99.4& 8& 2023\\ 
\cite{PhysRevLett.132.133603}&GaN & 60 & 4.3$\times$10$^5$&2.09$\times$10$^6$ & 5.8$\times$10$^8$& 108& 92.3& 7& 2024\\
\cite{PhysRevLett.133.083803}&SiN & - & 4.9$\times$10$^6$&3$\times$10$^7$ & 1.5$\times$10$^{11}$& 1438& 99.5& 1& 2024\\
This work&AlGaAs & 58.6 & 7.8$\times$10$^4$&6.5$\times$10$^7$ & 2.6$\times$10$^9$& 118& 93.1& 11& 2025\\ 
\bottomrule
\end{tabular}
}
\end{threeparttable}
\end{table}

\newpage

\end{document}